\def\year{2020}\relax
\documentclass[letterpaper]{article} 
\usepackage{aaai20}  
\usepackage{times}  
\usepackage{helvet} 
\usepackage{courier}  
\usepackage[hyphens]{url}  
\usepackage{graphicx} 
\usepackage{graphicx}  
\usepackage{amsmath} 
\usepackage{amssymb} 
\usepackage{multirow} 
\usepackage{pifont} 
\newcommand{\xmark}{\ding{55}} 
\newcommand{\cmark}{\ding{51}} 
\title{JSI-GAN: GAN-Based Joint Super-Resolution and Inverse \\Tone-Mapping with Pixel-Wise Task-Specific Filters for UHD HDR Video}
\author{\Large \textbf{Soo Ye Kim,\footnotemark[1]} \Large \textbf{Jihyong Oh,}\thanks{Both authors contributed equally to this work.} \Large \textbf{Munchurl Kim}\\ 
Korea Advanced Institute of Science and Technology\\
Republic of Korea\\
\{sooyekim, jhoh94, mkimee\}@kaist.ac.kr
}

\begin{document}

\maketitle

\begin{abstract}
Joint learning of super-resolution (SR) and inverse tone-mapping (ITM) has been explored recently, to convert legacy low resolution (LR) standard dynamic range (SDR) videos to high resolution (HR) high dynamic range (HDR) videos for the growing need of UHD HDR TV/broadcasting applications. However, previous CNN-based methods directly reconstruct the HR HDR frames from LR SDR frames, and are only trained with a simple L2 loss. In this paper, we take a \textit{divide-and-conquer} approach in designing a novel GAN-based joint SR-ITM network, called JSI-GAN, which is composed of three task-specific subnets: an image reconstruction subnet, a detail restoration (DR) subnet and a local contrast enhancement (LCE) subnet. We delicately design these subnets so that they are appropriately trained for the intended purpose, learning a pair of pixel-wise 1D separable filters via the DR subnet for detail restoration and a pixel-wise 2D local filter by the LCE subnet for contrast enhancement. Moreover, to train the JSI-GAN effectively, we propose a novel detail GAN loss alongside the conventional GAN loss, which helps enhancing both local details and contrasts to reconstruct high quality HR HDR results. When all subnets are jointly trained well, the predicted HR HDR results of higher quality are obtained with at least 0.41 dB gain in PSNR over those generated by the previous methods. The official Tensorflow code is available at \textit{https://github.com/JihyongOh/JSI-GAN}. 
\end{abstract}

\section{Introduction}

High dynamic range (HDR) videos can more realistically represent natural scenes, with higher bit depth per pixel and rich colors from a wider color gamut. They can be viewed on HDR TVs, which also tend to be UHD (Ultra High Definition) with very high resolution. However even with the latest UHD HDR TVs, the vast majority of the transmitted visual contents are still low resolution (LR), standard dynamic range (SDR) videos. There are also abundant existing legacy LR SDR videos, which brings about the need for appropriate conversion technologies that can generate high resolution (HR) HDR videos from LR SDR videos. 
\begin{figure} [ht]
\centering
\includegraphics[width=\columnwidth]{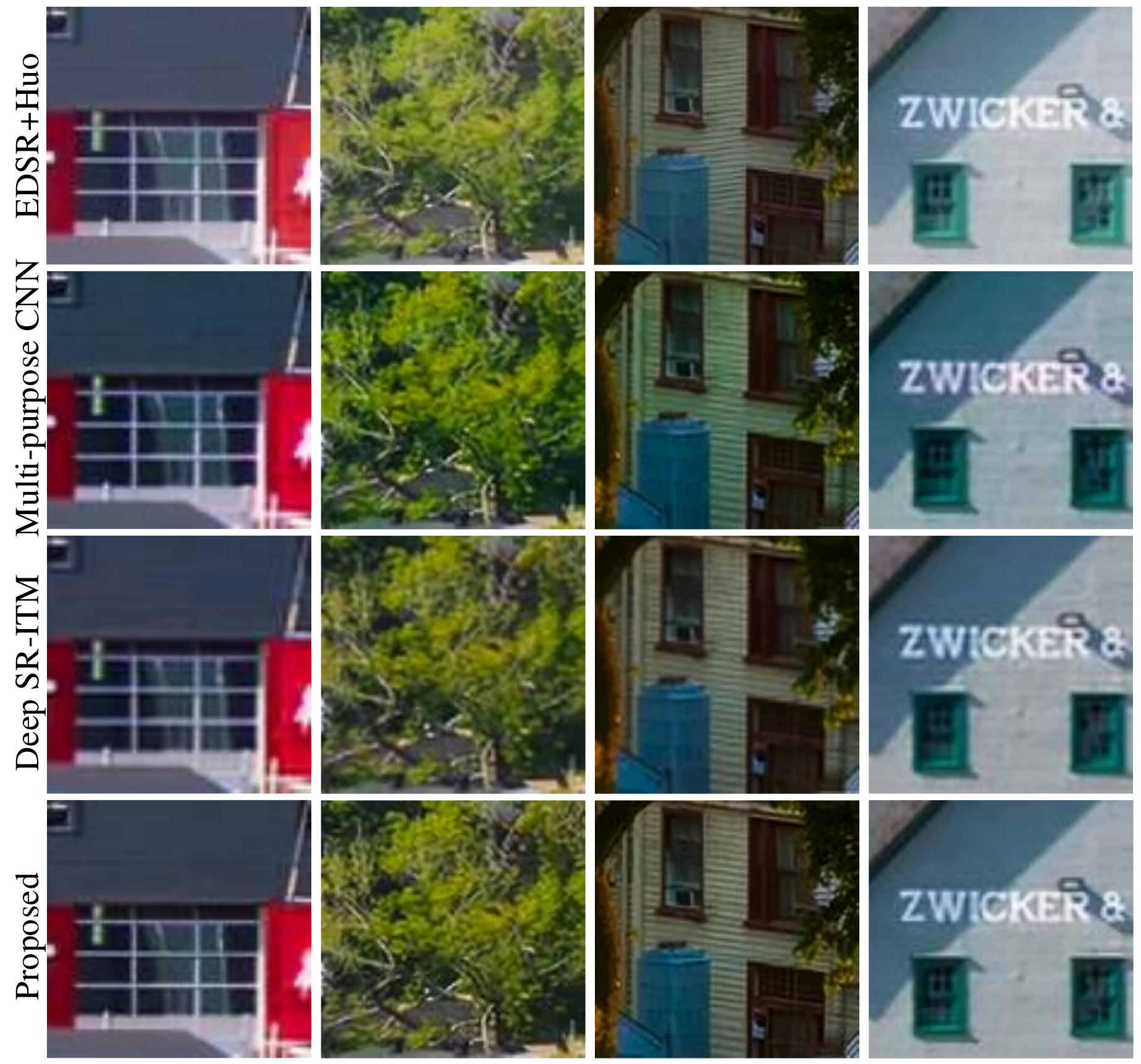}
\caption{Qualitative comparison against other methods. Our method can reconstruct fine lines with realistic textures.}
\label{fig:1}
\end{figure}
This can be achieved by joint super-resolution (SR) and inverse tone-mapping (ITM), where SR up-scales LR videos to HR, and ITM up-converts SDR videos to HDR.

In joint SR-ITM, it is important to restore details while up-scaling the image resolution, and to enhance local contrast while increasing the signal amplitudes. In this paper, we take a `\textit{divide-and-conquer}' approach by dividing this joint problem into three sub-tasks: image reconstruction (IR), detail restoration (DR), and local contrast enhancement (LCE). A single subnet is dedicated for each of the tasks, but all subnets are jointly trained for the same goal of joint SR-ITM. To overcome the limitations of conventional shared-convolution filters over input channels in each layer, we design a pair of pixel-wise 1D separable filters in the DR subnet for detail restoration and a pixel-wise 2D local filter in the LCE subnet for contrast enhancement. Moreover, the 1D separable and 2D local filters are designed to be scalable to up-scaling factors. This approach is inherently different from the approaches that directly produce the output HR HDR images. Furthermore, each input frame is divided into its base layer and a detail layer component by the guided filter \cite{he2012guided}. In our composite framework, the DR subnet, LCE subnet and the IR subnet are optimally combined to finally produce faithful HR HDR results.  

Generative adversarial networks (GANs) have been widely applied in low level vision tasks, such as SR, where they tend to generate images with high subjective (perceptual) quality but low objective quality (e.g. PSNR, SSIM, etc.). For joint SR-ITM, direct generation of output images based on the conventional GAN-based methods can lead to unsatisfying results with lack of details and unnatural local contrasts, since simultaneously enhancing the local contrast and details while increasing both the bit-depth and the spatial resolution becomes a very challenging task in training GAN-based frameworks. Therefore, our GAN-based joint SR-ITM method, called JSI-GAN, incorporates a novel detail loss that enforces its generator to mimic the perceptually realistic details in the ground truth, and a feature-matching loss that helps mitigate the drop in objective performance by stabilizing the training process.

\smallskip
Our contributions are summarized as follows:
\begin{itemize}
\item We first propose a GAN framework for joint SR-ITM, called JSI-GAN, with a novel detail loss and a feature-matching loss that enable the restoration of realistic details and force stable training.
\item The generator of JSI-GAN is designed to have task-specific subnets (DR/IR/LCE subnets) with pixel-wise 1D separable filters for local detail improvement and a 2D local filter for local contrast enhancement by considering local up-sampling operation given the up-scaling factor.
\item The DR subnet focuses on the high frequency components to elaborately restore the details of the HR HDR output, while the LCE subnet effectively restores the local contrast by focusing on the base layer component of the LR SDR input.
\end{itemize}

\section{Related Work}

\subsection{Pixel-Wise or Pixel-Aware Filter Learning}

In convolution layers, the same convolution kernels are applied on all spatial positions of the input, and once trained, the same kernels are used for any input image. Therefore, to consider pixel-wise and sample-wise diversity, Brabandere \textit{et al.} first introduced dynamic filter networks \cite{jia2016dynamic} in video and stereo prediction, where position-specific filters are predicted through a CNN and applied as an inner product on each pixel position of the input image. Since a different filter is applied to each pixel, and different filters are predicted from different input images, they allow for sample-specific and position-specific filtering. This operation is called \textit{dynamic local filtering}. 

This concept was successfully utilized in other video-related tasks, such as frame interpolation \cite{niklaus2017videob} and video SR \cite{jo2018deep}. Niklaus \textit{et al.}'s first idea \cite{niklaus2017videoa} was similar to that of \cite{jia2016dynamic} with 2D local filters being predicted, named as \textit{adaptive convolution}. Their extended work \cite{niklaus2017videob} with two 1D separable filters (horizontal/vertical) allowed to enlarge the final receptive field with the same number of parameters. For filter generation networks, the receptive field for the final output is solely defined by the size of the generated filter, implying that the depth or kernel sizes in the middle layers do not affect the final receptive field. Jo \textit{et al.} used 3D-CNNs and added an up-sampling feature in generating the 2D filters (dynamic up-sampling filters), while incorporating a conventional residual network with direct reconstruction.

In our architecture, we design (i) the DR subnet with pixel-wise 1D separable horizontal and vertical filters to capture the \textit{distinct details}; (ii) the LCE subnet with pixel-wise 2D local filters so that it can focus on the \textit{local region contrast}. With the same number of filter parameters, 1D separable filters are coarse representations of a larger region, whereas 2D filters are finer representations of a local receptive area. In our framework, the DR subnet, LCE subnet and the IR subnet are combined to produce the final HR HDR images. 
\begin{figure*}
\centering
\includegraphics[scale=0.8]{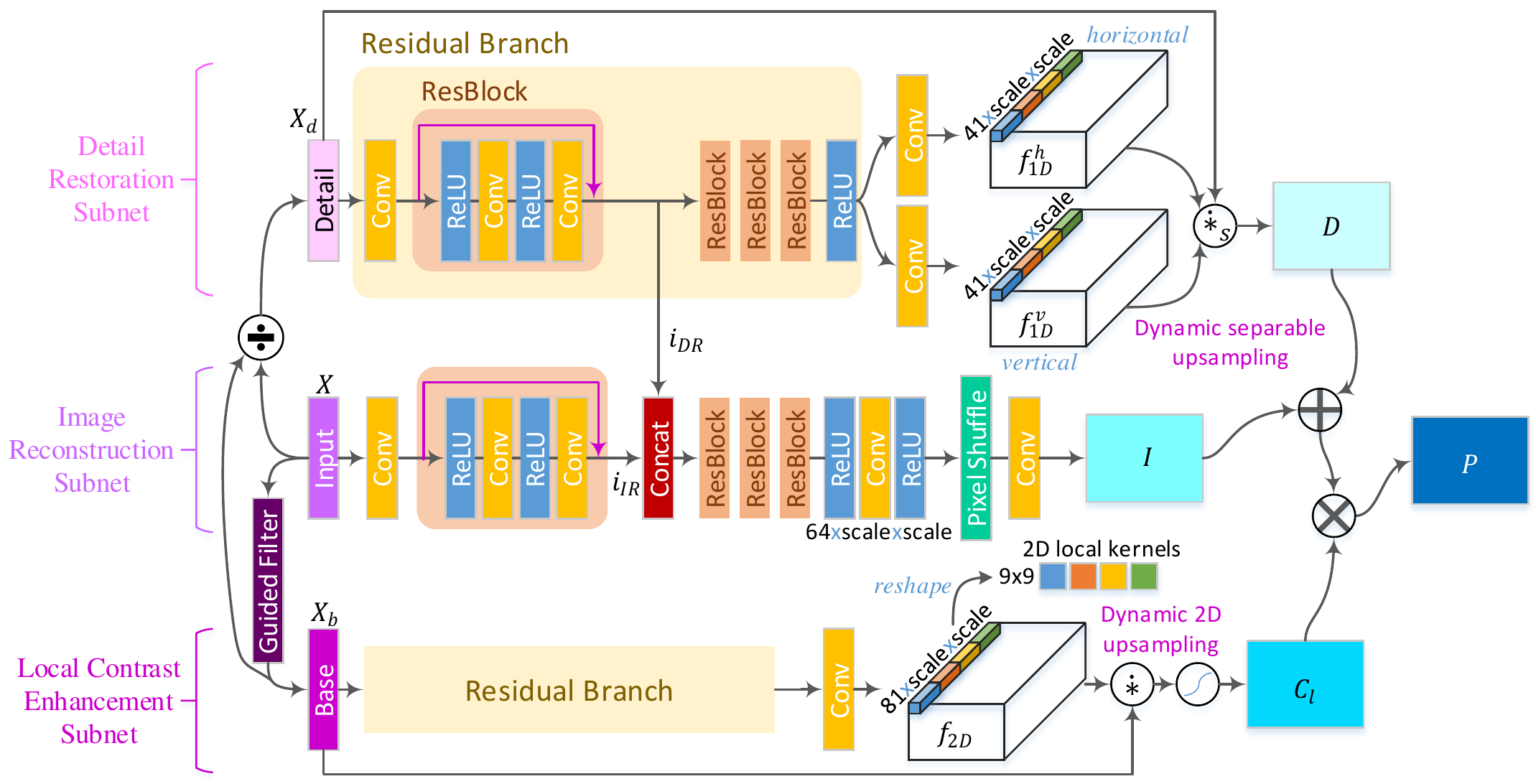}
\caption{Network architecture of JSInet (Generator).}
\label{fig:2}
\end{figure*}

\subsection{Generative Adversarial Networks}

A GAN-based framework is typically composed of a generator and a discriminator, which are trained in an adversarial way, to force the generator to synthesize realistic images that are indistinguishable by the discriminator \cite{goodfellow2014generative}. Many advanced techniques and variants of GANs have achieved significant performance improvements in various computer vision tasks, especially in image restoration, such as image dehazing \cite{qu2019enhanced}, SR \cite{wang2018esrgan,ledig2017photo}, denoising \cite{chen2018image} and image enhancement \cite{chen2018deep}. These methods adopted various GAN-based frameworks to improve the perceptual quality for their individual purposes, and they generally train the main network as the generator, which is first trained with a pixel-wise norm-based loss, and is then fine-tuned by introducing an adversarial loss with the discriminator.

Motivated by the enhanced perceptual quality of GAN-based methods, we also design a GAN-based framework for joint SR-ITM. However, simply adopting conventional GANs for joint SR-ITM leads to difficulty in training the network for a more complex task of jointly improving the local variations of contrast and details along with up-converting to both higher bit depth per pixel and larger spatial resolution. To guide the generator in a more effective way to fool the discriminator and generate perceptually realistic HR HDR results with the GAN-based framework, we propose a novel detail GAN loss, which jointly optimizes a second discriminator for the detail components decomposed from the HR HDR prediction and the ground truth. We also employ the feature-matching loss that is measured in the feature space of the discriminator to reduce the drop in objective performance for a more stable training. 

\subsection{Joint SR-ITM}

Deep-learning-based joint SR-ITM is a recent topic rising with the advent of UHD HDR TVs. The first CNN-based joint SR-ITM method \cite{kim2018multi} took a multi-task learning approach and focused on the advantages of performing the individual SR and ITM tasks simultaneously, along with the joint SR-ITM task. The more recent method is the Deep SR-ITM \cite{kim2019deep}, where the input is decomposed, and element-wise modulations are inserted to introduce spatially-variant operations. We also use the decomposed input for the DR subnet, but our architecture incorporates pixel-wise filters of two kinds: separable 1D kernels and 2D local kernels with up-sampling that are used to generate the final HR HDR output by filtering operations. Moreover, our network is a GAN-based framework, differing from the previous methods.

\section{Proposed Method}

We propose a GAN-based framework for joint SR-ITM, called JSI-GAN, where the generator, JSInet, is composed of three different task-specific subnets.

\subsection{Network Architecture (Generator)}
In joint SR-ITM, it is important to restore the high frequency details and enhance local contrast with the increase in image resolutions and pixel amplitudes to generate a high quality HR HDR image. Therefore, we design three subnets dedicated for each of these subtasks as a \textit{divide-and-conquer} approach: the image reconstruction (IR) subnet reconstructs a coarse HR HDR image; the detail restoration (DR) subnet restores the details to be added on the coarse image; and the local contrast enhancement (LCE) subnet generates a local contrast mask to boost the contrast in this image. A detailed structure of JSInet is depicted in Fig. \ref{fig:2}.

\subsubsection{Detail Restoration (DR) Subnet}
For the DR subnet, the input is the detail layer $X_{d}$, containing the high frequency components of the LR SDR input image $X$. $X_{d}$ is given by,
\begin{align} 
	X_{d} = X\oslash X_{b},
\label{eq1}
\end{align}
where $X_{b}$ is the guided filtered output of $X$, and $\oslash$ denotes element-wise multiplication, as in \cite{kim2019deep}. In our implementations, a small value $10^{-15}$ is added to the denominator to prevent $X_{d}$ from diverging in case $X_{b}$ approaches zero. $X_{d}$ is utilized to generate horizontal and vertical 1D separable filters. Formally, a residual block (ResBlock) RB is defined as,
\begin{align}
    RB(x) = (Conv\circ RL\circ Conv\circ RL)(x)+x,
\label{eq2}
\end{align}
where $x$ is the input to the ResBlock, $Conv$ is a convolution layer, and $RL$ is a ReLU activation. Then, the horizontal 1D filter $f^{h}_{1D}$ is obtained by,
\begin{align}
    f^{h}_{1D} = (Conv\circ RL\circ RB^{4}\circ Conv)(X_{d}),
\label{eq3}
\end{align}
where $RB^{n}$ denotes a serial cascade of $n$ ResBlocks. The vertical 1D filter $f^{v}_{1D}$ can be obtained in the same way as Eq. (\ref{eq3}). As shown in Fig. \ref{fig:2}, all layers except the last convolution layer are shared when producing $f^{h}_{1D}$ and $f^{v}_{1D}$.

Each of the two last convolution layers consists of $41\times scale\times scale$ output channels, where 41 is the length of the 1D separable kernel, each applied onto its corresponding grid location, and $scale\times scale$ takes into account the pixel shuffling operation for the up-scaling factor $scale$. Hence, this dynamic separable up-sampling operation ($\dot{*}_{s}$) applies two 1D separable filters while producing the spatially-upscaled output. Then, the final filtered output of the DR subnet is given by,
\begin{align}
    D = X_{d}\:\dot{*}_{s}\:(f^{v}_{1D}, f^{h}_{1D}).
\label{eq4}
\end{align}
The generated 1D kernels are position-specific, and also detail-specific, as different kernels are generated for different detail layers, unlike convolution filters that are fixed after training. In our implementations, $f^{v}_{1D}$ was first applied to the detail layer via local filtering for each scale channel, followed by applying $f^{h}_{1D}$ on its output. Finally, pixel shuffle was applied on the final filtered output with $scale\times scale$ channels for spatial up-scaling. With the same number of parameters, the 1D separable kernels represent a wider receptive field ($2k$ parameters), but as a coarse approximation compared to a 2D kernel ($k^{2}$ parameters). In our case, the separable 1D kernels of size $k=41$ can be compared to the 2D kernel of size $k=9$, with similar number of parameters.

\subsubsection{Local Contrast Enhancement (LCE) Subnet}
The base layer $X_{b}$, obtained through guided filtering, is used as the input to the LCE subnet. The LCE subnet generates a $9\times 9$ 2D local filter at each pixel grid position. As in the DR subnet, it is also an up-sampling filter, and has $9\times 9\times scale\times scale$ output channels in the last layer. The 2D filter $f_{2D}$ is then given by,
\begin{align}
    f_{2D} = (Conv\circ RL\circ RB^{4}\circ Conv)(X_{b}).
\label{eq5}
\end{align}
With the 2D local filter predicted, dynamic 2D up-sampling operation ($\dot{*}$) is performed, and $2\times sigmoid(x)$ is applied so that the the output lies in the range [0, 2] with the middle value at $x=0$ being 1. Formally, the final output $C_{l}$ of the LCE subnet is given as,
\begin{align}
    C_{l} = 2\times sigmoid(X_{b}\:\dot{*}\:f_{2D}).
\label{eq6}
\end{align}
As $C_{l}$ is considered an LCE mask, and is element-wisely \textit{multiplied} onto the sum of the two outputs from the IR and DR subnets, the JSInet converges better with the scaled sigmoid function. Without it, the pixel values of the initial predicted outputs (after multiplying with $C_{l}$) are too small, entailing much longer training time for the final HR HDR output of JSInet to reach an appropriate pixel range.

\subsubsection{Image Reconstruction Subnet}
For the IR subnet, the LR SDR input $X$ as itself is entered, to first produce the intermediate features $i_{IR}$ as shown in Fig. \ref{fig:2}, given by,
\begin{align}
    i_{IR} = (RB\circ Conv)(X).
\label{eq7}
\end{align}
Then, $i_{IR}$ is concatenated with $i_{DR}$ from the DR subnet, and the final output $I$ of the IR subnet is directly generated (without local filtering) as,
\begin{multline}
I = (Conv\circ PS\circ RL\circ Conv\circ RL\circ RB^{3})([i_{IR},i_{DR}]),
\label{eq8}
\end{multline}
where $PS$ is a pixel-shuffle operator and $[x, y]$ is the concatenation of $x$ and $y$ in the channel direction. 

Then, the final HR HDR prediction $P$ is generated by adding the details ($D$) to $I$ followed by multiplying the contrast mask ($C_{l}$) to the result, given by,
\begin{align}
    P = (I+D)\times C_{l}.
\label{eq9}
\end{align}
We provide an ablation study on the three subnets and demonstrate that they are acting according to their given tasks in the later sections of the paper.
\begin{figure}
\centering
\includegraphics[scale=0.8]{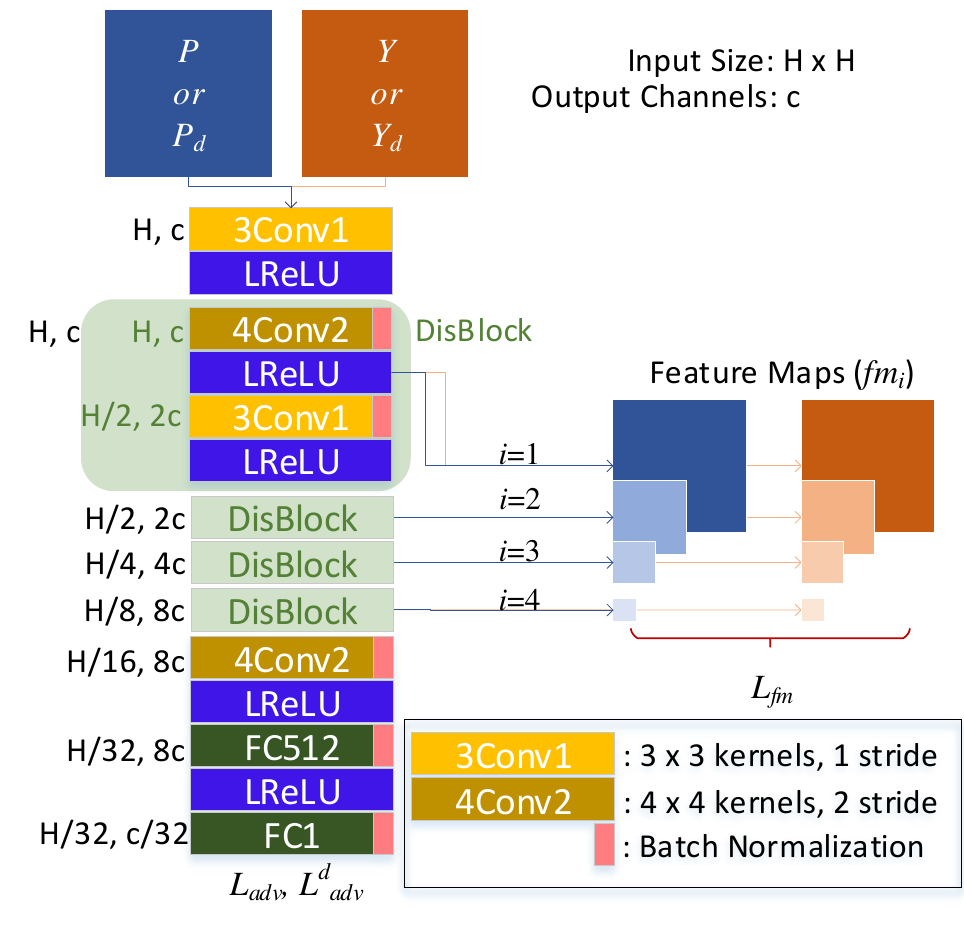}
\caption{Discriminator architecture of JSI-GAN.}
\label{fig:3}
\end{figure}

\subsection{GAN-based Framework}
\subsubsection{Discriminator}
We employ a GAN-based framework, where the discriminator is designed as shown in Fig. \ref{fig:3} with spectral normalization \cite{miyato2018spectral} for stable learning. The discriminator distinguishes the predicted HR HDR image ($P$) generated by the generator and the ground truth image ($Y$) alternatively during training. As shown in Fig. \ref{fig:3}, when the input $x$ ($P$ or $Y$) is given to the discriminator $D^{f}$, the output is obtained by,
 \begin{multline}
    D^{f}(x) = (BN \circ FC1 \circ BN \circ FC512 \circ LRL \\ \circ BN \circ 4Conv2 \circ DB^{4} \circ LRL\circ 3Conv1)(x).
\label{eq10}
\end{multline}
where $BN$ is batch normalization \cite{ioffe2015batch}, $LRL$ is a Leaky ReLU (LReLU) activation \cite{maas2013rectifier} with a slope size of 0.2, FC$k$ is a fully-connected layer with $k$ output channels, and $kConvs$ denotes $k\times k$ kernels with a stride $s$. The input sizes (H $\times$ H) and output channels (c) for each layer are specified as (H, c) on the left of each layer/block in Fig. \ref{fig:3}. $DB^{n}$ denotes $n$-times serially connected DisBlocks, where $DB$ (DisBlock) is given as,
\begin{multline}
    DB(x) = (LRL\circ BN \circ 3Conv1 \circ  LRL \\\circ BN \circ 4Conv2)(x).
\label{eq11}
\end{multline}

\subsubsection{Adversarial Loss}
The RaHinge GAN loss \cite{jolicoeur2018relativistic} is adopted as a basic adversarial loss for efficient training, which is given by
\begin{multline}
    L_{adv}^{D}=\mathop{\mathbb{E}}_{Y}[max(0,\tilde{Q}^{(-)}_{Y,P})]+\mathop{\mathbb{E}}_{P}[max(0,\tilde{Q}^{(+)}_{P,Y})]
\label{eq12}
\end{multline}
\begin{multline}
    L_{adv}^{G}=\mathop{\mathbb{E}}_{P}[max(0,\tilde{Q}^{(-)}_{P,Y}))]+\mathop{\mathbb{E}}_{Y}[max(0,\tilde{Q}^{(+)}_{Y,P}))]
\label{eq13}
\end{multline}
where $L_{adv}^{D}$ and $L_{adv}^{G}$ denote the RaHinge GAN losses for the discriminator $D^{f}$ and the generator $G$, respectively, and $\tilde{Q}^{(\pm)}_{P,Y}=1\pm\tilde{D}_{P,Y}$ with $\tilde{D}_{P,Y}=D_{f}(P)-\mathop{\mathbb{E}}_{Y}D_{f}(Y)$. It should be noted in Eq. (\ref{eq13}) that $L_{adv}^{G}$ contains both the $Y$ and $P$, meaning that the generator is trained by gradients from both the $Y$ and $P$ during the adversarial training. We also use a feature-matching loss $L_{fm}=\sum_{i=1}^{4}{\lVert fm_{i}(Y)-fm_{i}(P)\rVert_{2}}$, where the L2 loss is measured between feature maps $fm_{i}(\cdot)$ of $Y$ and $P$, from the first LReLU output of the $i$-$th$ DisBlock as shown in Fig. \ref{fig:3}. However, simple utilization of the above two losses is insufficient to effectively train the generator for joint SR-ITM. 

\subsubsection{Detail GAN Loss}

We propose a novel detail GAN loss $L_{adv}^{d}$ for the joint SR-ITM task, in order to enforce more stable training and generate visually pleasing HR HDR results. $L_{adv}^{d}$ is calculated according to Eq. (\ref{eq12}) and Eq. (\ref{eq13}) by replacing $Y$ with $Y_{d}$ and $P$ with $P_{d}$, where the subscript $d$ denotes the detail layer component of the original image. For $L_{adv}^{d}$, we adopt a second discriminator ($D_{2}$) distinguished from the first discriminator ($D_{1}$), both of which have the same architecture as shown in Fig. \ref{fig:3}, but $D_{2}$ alternatively takes two inputs $P_{d}$ and $Y_{d}$, calculated by Eq. (\ref{eq1}). $L_{adv}^{d}$ not only guides the generator for a more stable training but also helps improve both local contrasts and details in the predicted HR HDR images. 
\subsubsection{Total Loss}
 
The total loss for our proposed GAN-based framework for joint SR-ITM is given by
 \begin{align}
    L_{D_{1}}=L_{adv}^{D_{1}}, \;L_{D_{2}}=\lambda_{d}\cdot L_{adv}^{d, D_{2}},
\label{eq14}
\end{align}

 \begin{multline}
    L_{G}=\lambda_{rec}\cdot \lVert Y-P\rVert_{2}+\lambda_{adv}\cdot (L_{adv}^{G}+\lambda_{d}\cdot L_{adv}^{d, G})\\+\lambda_{fm}\cdot (L_{fm}+\lambda_{d}\cdot L_{fm}^{d}),
\label{eq15}
\end{multline}
where the superscript $d$ means the loss for detail layer components ($P_{d}$, $Y_{d}$). We provide another ablation study on the losses $L_{fm}$ and $L^{d}$, and especially show the effect of the newly proposed $L_{adv}^{d}$, in the later sections of the paper.

\section{Experiment Results}

\subsection{Experiment Conditions}

For training, the generator was first pre-trained based only upon the L2 loss with the initial learning rate of $10^{-4}$, which is then decreased by a factor of 10 at epochs 200 and 225, of total 250 epochs, yielding the JSInet. Then it was fine-tuned based on a stable GAN-based framework with three losses ($L_{G}$, $L_{D_{1}}$, $L_{D_{2}}$) at the initial learning rate of $10^{-6}$ that linearly decays to zero from the 5-\textit{th} epoch of total 10 epochs, which finally yields the JSI-GAN. For training, we used three Adam optimizers \cite{kingma2014adam} for minimizing $L_{D_{1}}$, $L_{D_{2}}$ and $L_{G}$, and all convolution weights were initialized by the Xavier method \cite{glorot2010understanding}. The generator $G$ and the two discriminators ($D_{1}$, $D_{2}$) are trained alternatively by the three corresponding Adam optimizers. The weighting parameters for the losses were empirically set to $\lambda_{rec}=1$, $\lambda_{adv}=1$, $\lambda_{fm}=0.5$ and $\lambda_{d}=0.5$. 

In the generator, the kernel size of the convolution filters were set to $3\times3$ with 64 output channels, except for the last layer that predicts local filters for the DR subnet and the LCE subnet, and the layer before pixel shuffle for the IR subnet. The structure of both $D_{1}$ and $D_{2}$ has the channel outputs of $c=32$. The LR SDR patches of size $80\times 80$ $(H=80)$ or $40\times 40$ $(H=40)$ were cropped from 8-bit YUV frames in BT.709 \cite{bt709} for scale factor 2 and 4 respectively, and the ground truth (HR HDR) patches of size $160\times 160$ $(H=160)$ were cropped from the corresponding 10-bit YUV frames in BT.2020 \cite{bt2020} color container after PQ-OETF \cite{st2084}, following the setting in previous work \cite{kim2019deep}. For training and testing the JSI-GAN, we utilized the 4K HDR dataset used in \cite{kim2019deep} and adopted Tensorflow 1.13 in our implementations.

\begin{table}
\begin{center}
\caption{Ablation study on the subnets.}
\label{table:1}
\scalebox{0.85}{
\begin{tabular}{ c|c|c|c|c|c|c }
\hline\hline
Variant & (a) & (b) & (c) & (d) & (e) & (f) \\
\hline\hline
IR & \cmark & \cmark & \cmark & \cmark & \cmark & \cmark\\
\hline
DR & \xmark & \cmark (-) & \cmark & \xmark & \cmark (-) & \cmark\\
\hline
LCE & \xmark & \xmark & \xmark & \cmark & \cmark & \cmark\\
\hline\hline
PSNR & 35.44 & 35.89 & 35.88 & 35.67 & 35.68 & \textbf{35.99}\\
\hline
SSIM & 0.9756 & 0.9762 & \textbf{0.9773} & 0.9761 & 0.9756 & 0.9768\\
\hline\hline
\multicolumn{7}{l}{*(-): Subtraction instead of division to obtain the detail layer.}\\
\multicolumn{7}{l}{**PSNR is measured in dB.}
\end{tabular}}
\end{center}
\end{table}

\subsection{Performance of JSInet}
We first investigate the performance of JSInet that is trained solely on the L2 loss \textit{without} the GAN framework, by analyzing the efficacy of its three subnets.

\subsubsection{Ablation Study of Subnets}
We first performed an ablation study on the three subnets in JSInet, by retraining different variants of the network. Table \ref{table:1} shows the PSNR/SSIM performance of six combinations (variants) of the three subnets, where the IR subnet is essential for all cases. As shown in column (c) of Table \ref{table:1}, employing the DR subnet to the IR subnet brings 0.44 dB gain in PSNR, and additionally using the LCE subnet in column (f), further increases the PSNR by 0.11 dB, resulting in a total 0.55 dB gain over only using the IR subnet in column (a). It is also noted that the LCE subnet brings a higher performance gain when fused with the IR subnet (column (d)) with 0.23 dB gain, meaning that the LCE subnet is complementarily beneficial with the DR subnet. 

We have also provided the experiment results for a different decomposition strategy, subtraction, on the input images instead of division in Eq. (\ref{eq1}), with the sign (-) in Table \ref{table:1}. For the structure in column (b) and (c), there is minimal difference in PSNR, although division yields slightly better performance for SSIM. However, when the LCE subnet is jointly utilized, division outperforms subtraction by a larger margin of 0.31 dB in PSNR, since for column (e), as the local contrast map $C_{l}$ (obtained from the base layer) is multiplied to the coarse image $I$ with the added details $D$ in Eq. (\ref{eq9}), the subtraction operation is unmatched with the multiplication operation.

Some additional results not marked in Table \ref{table:1} were the following. (i) If the concatenation of $i_{DR}$ was removed, there was a 0.02 dB drop in PSNR from the network in Table \ref{table:1} (c). (ii) If feature maps $i_{LCE}$ from the LCE subnet were additionally concatenated along with $i_{IR}$ and $i_{DR}$, there was a 0.25 dB drop in PSNR from the network in Table \ref{table:1} (f). Note that only $i_{IR}$ and $i_{DR}$ are stacked in the final generator architecture of our proposed JSInet in Fig. \ref{fig:2}. (iii) If the image $X$ was stacked with $X_{d}$ for the input of the DR subnet, following the idea of providing guidance to the detail layer in \cite{kim2019deep}, there was actually a minimal decrease, or no difference in PSNR, when experimented with the network in Table \ref{table:1} (b) and Table \ref{table:1} (c), respectively.

\begin{figure}
\centering
\includegraphics[width=\columnwidth]{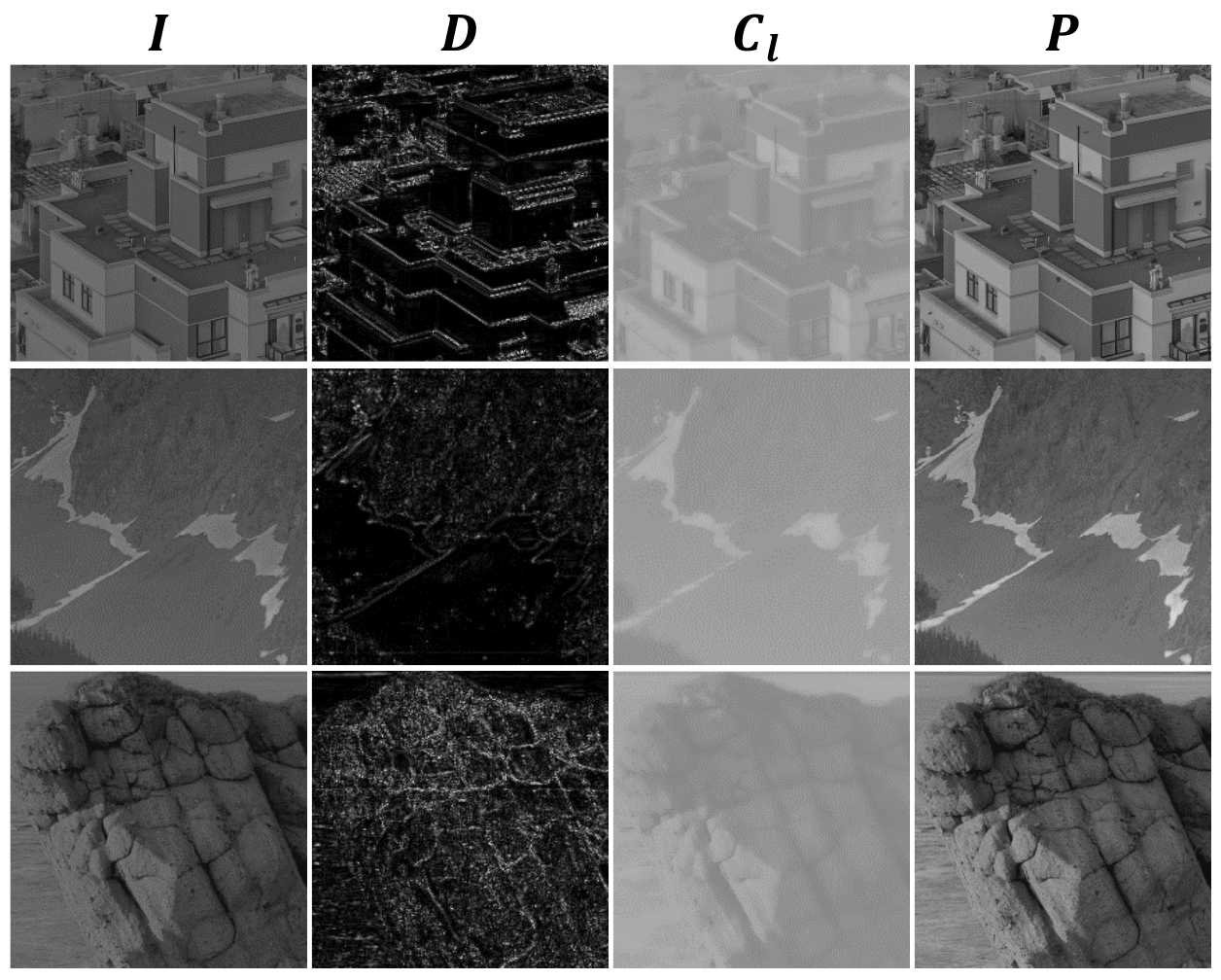}
\caption{Visualization of the subnet predictions.}
\label{fig:6}
\end{figure}

\begin{table}
\begin{center}
\caption{Combinations of direct reconstruction and filter generation for the network with IR/DR subnets.}
\label{table:5}
\scalebox{0.9}{
\begin{tabular}{ c|c|c|c }
\hline\hline
IR subnet & direct & filter & direct\\
\hline
DR subnet & direct & direct & filter\\
\hline\hline
PSNR (dB) & 35.73 & 34.14 & \textbf{35.88} \\
\hline
SSIM & 0.9765 & 0.9662 & \textbf{0.9773}\\
\hline\hline
\end{tabular}}
\end{center}
\end{table}

\subsubsection{Visual Analysis}
To verify that each of the subnets is focusing on their dedicated tasks, the intermediate predictions ($I, D, C_{l}$) of the subnets and the final prediction $P$ are visualized in Fig. \ref{fig:6}. For visualization, $I, |D|, C_{l}$ and $P$ were first linearly scaled to a maximum value of 8 bits/pixel, and $|D|$ was further scaled by 64 for better visualization. In Fig. \ref{fig:6}, the added details $D$ are invariant to the brightness context (1st and 2nd rows) and lighting conditions (3rd row), focusing only on the edges and texture as intended. The LCE mask $C_{l}$ effectively modulates the local contrast, producing the final $P$ with enhanced contrast.

\begin{figure} [t]
\centering
\includegraphics[width=\columnwidth]{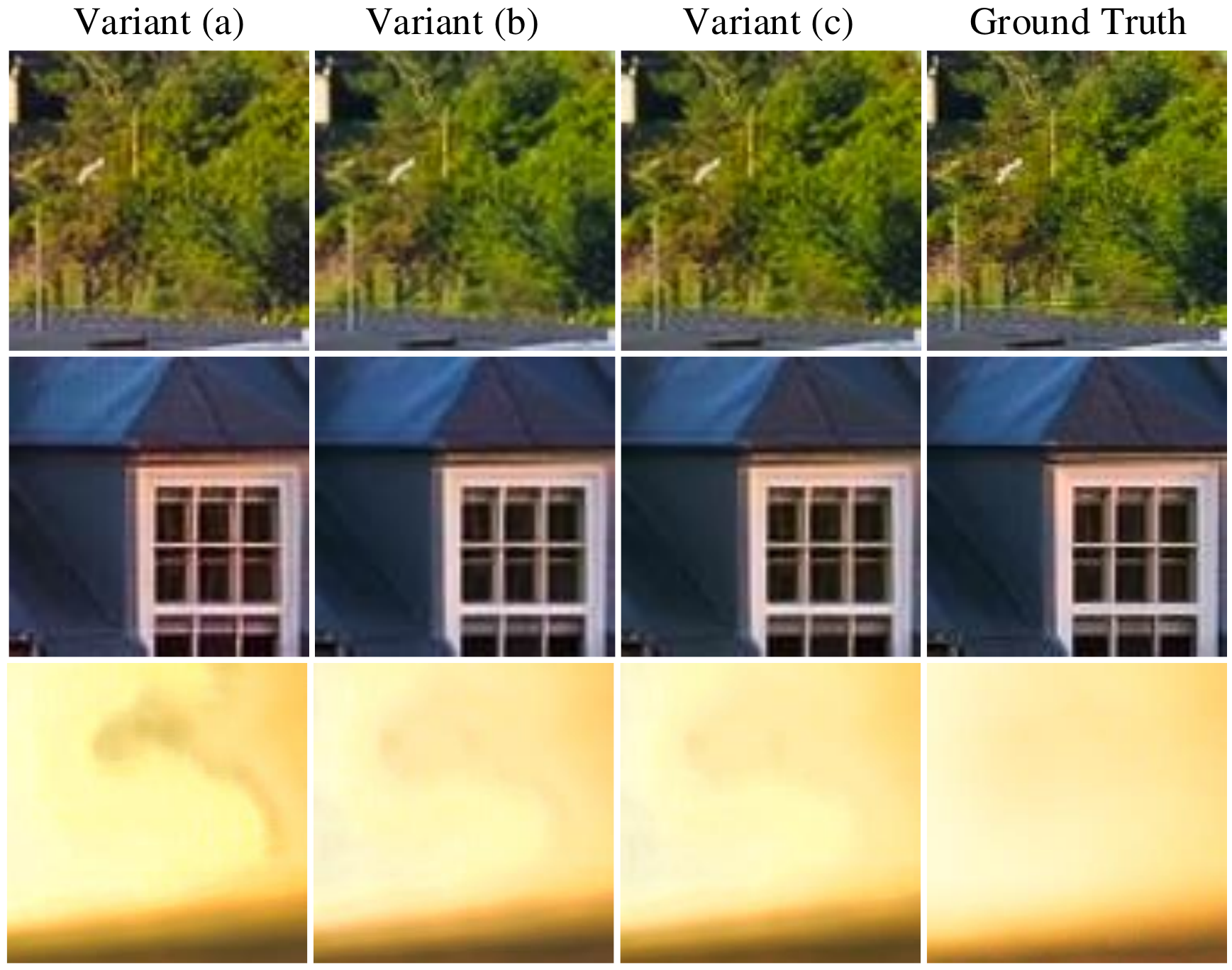}
\caption{Effect of the weight parameters $\lambda_{fm}$ and $\lambda_{d}$.}
\label{fig:4}
\end{figure}

\begin{table} [t]
\begin{center}
\caption{Ablation study on the losses.}
\label{table:4}
\scalebox{0.8}{
\begin{tabular}{ c|c|c|c|c|c|c }
\hline\hline
Scale & \multicolumn{3}{c|}{$\times$2} & \multicolumn{3}{c}{$\times$4}\\
\hline
Variant & (a) & (b) & (c) & (d) & (e) & (f)\\
\hline\hline
$\lambda_{fm}$ & \xmark & \cmark & \cmark & \xmark & \cmark & \cmark \\
\hline
$\lambda_{d}$ & \xmark & \xmark & \cmark & \xmark & \xmark & \cmark\\
\hline\hline
PSNR (dB) & 33.73 & 35.63 & \textbf{35.73} & 32.11 & 33.48 & \textbf{33.50}\\
\hline
SSIM & 0.9643 & 0.9754 & \textbf{0.9763} & 0.9478 & \textbf{0.9577} & 0.9572\\
\hline\hline
\end{tabular}}
\end{center}
\end{table}

\subsubsection{Filter Generation Module}
To analyze the benefits of generating local filters, we performed an experiment with the combinations of direct reconstruction and kernel generation for a simple network with only the IR and DR subnet, as shown in Table \ref{table:5}. For the DR subnet, it is important to apply location-varying filters, as uniform convolution operations would blur the high frequency details. If the detail map $D$ is directly generated, there is a performance drop of 0.15 dB in PSNR, as shown in Table \ref{table:5}. Moreover, if the filter generation module is used for the IR subnet, there is a drastic drop in PSNR performance compared to the proposed combination in the last column. For the LCE subnet, we employ a 2D filter generation module so that local information can be considered. If 1D separable kernels were predicted instead for the LCE subnet, there was a 0.07 dB drop in performance gain.

\begin{table*}
\begin{center}
\caption{Quantitative Comparison.}
\label{table:3}
\scalebox{0.9}{
\begin{tabular}{ l|c|c|c|c|c|c||c|c }
\hline\hline
Method & Scale & PSNR (dB) & SSIM & mPSNR (dB) & MS-SSIM & HDR-VDP (Q) & Inf. Time [s] & No. Parameters \\
\hline\hline
EDSR+Huo \textit{et al.} & $\times2$ & 29.76 & 0.8934 & 31.81 & 0.9764 & 58.95 & - & -\\
EDSR+Eilertsen \textit{et al.} & $\times2$ & 25.80 & 0.7586 & 28.22 & 0.9635 & 53.51 & - & -\\
Multi-purpose CNN & $\times2$ & 34.11 & 0.9671 & 36.38 & 0.9817 & 60.91 & 0.49 & 250K \\
Deep SR-ITM & $\times2$ & 35.58 & 0.9746 & 37.80 & 0.9839 & \textbf{61.39} & 5.02 & 2.50M \\
\hline\hline
\textbf{JSInet} (w/o GAN)& $\times2$ & \textbf{35.99} & \textbf{0.9768} & \textbf{38.20} & \textbf{0.9843} & 60.58 & 1.72 & 1.45M \\
\textbf{JSI-GAN} (w/ GAN)& $\times2$ & 35.73 & 0.9763 & 37.96 & 0.9841 & 60.80 & 1.72 & 1.45M \\
\hline\hline
EDSR+Huo \textit{et al.} & $\times4$ & 28.90 & 0.8753 & 30.92 & 0.9693 & 55.59 & - & -\\
EDSR+Eilertsen \textit{et al.} & $\times4$ & 26.54 & 0.7822 & 28.75 & 0.9631 & 53.88 & - & -\\
Multi-purpose CNN & $\times4$ & 33.10 & 0.9499 & 35.26 & 0.9758 & 56.41 & 0.34 & 283K\\
Deep SR-ITM & $\times4$ & 33.61 & 0.9561 & 35.73 & 0.9748 & 56.07 & 4.06 & 2.64M\\
\hline\hline
\textbf{JSInet} (w/o GAN) & $\times4$ & \textbf{33.74} & \textbf{0.9598} & \textbf{35.93} & \textbf{0.9759} & \textbf{56.45} & 1.05 & 3.03M \\
\textbf{JSI-GAN} (w/ GAN)& $\times4$ & 33.50 & 0.9572 & 34.82 & 0.9743 & 56.41 & 1.05 & 3.03M \\
\hline\hline
\end{tabular}}
\end{center}
\end{table*}

\begin{figure}
\centering
\includegraphics[width=\columnwidth]{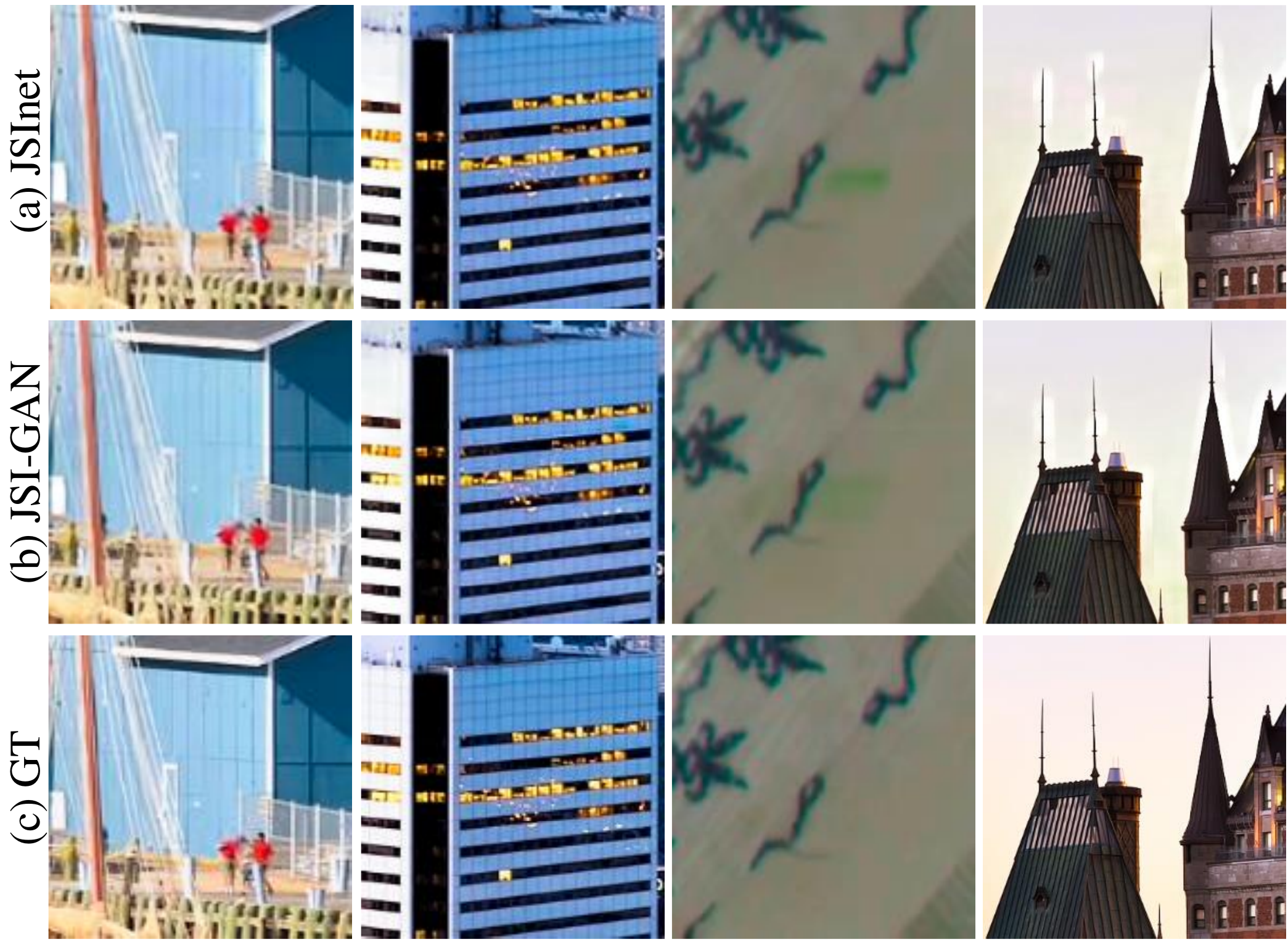}
\caption{Visual comparison: JSInet and JSI-GAN.}
\label{fig:7}
\end{figure}

\subsection{Performance of JSI-GAN}
We also conducted an ablation study on the efficacy of various losses in terms of weighting parameters $\lambda_{fm}$ and $\lambda_{d}$, to investigate their effects. Table \ref{table:4} shows the average PSNR (dB) and SSIM performance of the JSI-GAN variants, each of which was trained via combinations with/without $\lambda_{fm}$ and $\lambda_{d}$, for scales $\times2$ and $\times4$. If JSI-GAN is only trained with the basic adversarial loss where $\lambda_{rec}=1$, $\lambda_{adv}=1$, $\lambda_{fm}=0$ and $\lambda_{d}=0$, severe performance degradation is observed with at most 2 dB drop in PSNR (variants (a) and (d)). By comparing variants (b) to (a) and (e) to (d), additionally adopting the feature-matching loss $L_{fm}$ between \textit{P} and \textit{Y} helps significantly in improving objective performance, with 1.9dB and 1.37dB gain in PSNR, respectively. Finally, our proposed detail GAN loss $L_{adv}^{d}$ between $P_{d}$ and $Y_{d}$ allows for further improvements in the quantitative performance as shown in variants (c) and (f) of Table \ref{table:4}. 

The effect of the losses is also shown qualitatively in Fig. \ref{fig:4}. Just simply adopting the basic GAN loss not only degrades the quantitative performance, but also severely deteriorates the visual qualities with the checkerboard artifacts and unnatural details/contrasts, as shown in the leftmost column in Fig. \ref{fig:4}. Although the feature-matching loss helps the generator improve the overall visual quality when comparing the variant (b) to (a), the proposed detail-component-related losses ($L_{adv}^{d, D}$, $L_{adv}^{d, G}$, $L_{fm}^{d}$) additionally improve both visual qualities and objective performances comparing the variant (c) to the variants (a) and (b). As a result, the final JSI-GAN enables the restoration of realistic details with stable training while allowing for high objective quality of the HR HDR predictions. 

\subsection{Visual Comparison: JSInet and JSI-GAN}
We also compare the visual results of the JSInet (without GAN) and JSI-GAN (with GAN). Although there is a slight drop in the performance of JSI-GAN in terms of PSNR (dB) in Table \ref{table:3}, there are some qualitative differences as shown in Fig. \ref{fig:7}. Since JSInet is only trained via reconstruction loss (L2) \textit{without} the GAN framework, the visual results of JSInet shown in the 1st row of Fig. \ref{fig:7} tends to be blurred in general. By comparing the row (b) to row (a) in 1st and 2nd column in Fig. \ref{fig:7}, both local details and contrasts are boosted by utilizing proposed GAN-based framework. Furthermore, the ringing artifacts or color artifacts observed in the homogeneous regions (especially in background) of the results by JSInet, shown in 3rd and 4th columns of Fig. \ref{fig:7}, are also enhanced by JSI-GAN. 

\subsection{Comparison with Other Methods}
We compare our JSI-GAN with the two previous joint SR-ITM methods, the Multi-purpose CNN \cite{kim2018multi} and Deep SR-ITM \cite{kim2019deep}. The proposed JSI-GAN is trained on the same data as Deep SR-ITM, and the Multi-purpose CNN was re-trained on the same data. Additionally, the JSI-GAN is also compared with the cascades of an SR method, EDSR \cite{lim2017enhanced}, and two ITM methods \cite{huo2014physiological,eilertsen2017hdr}. Note that Huo \textit{et al}l's method is not data-driven, and thus invariant to training data. Since Eilertsen \textit{et al}.'s method was designed to be trained only for saturated pixel regions, it was not trained with our data that does not contain enough saturated pixel regions (unlike luminance-map-type data used in their dataset). The pre-trained model of EDSR as provided in the official code is used. The previous methods were compared following the experiment protocol on the ITM prediction pipeline and visualization as described in Deep SR-ITM \cite{kim2019deep}.

\begin{figure*}[t]
\centering
\includegraphics[width=\textwidth]{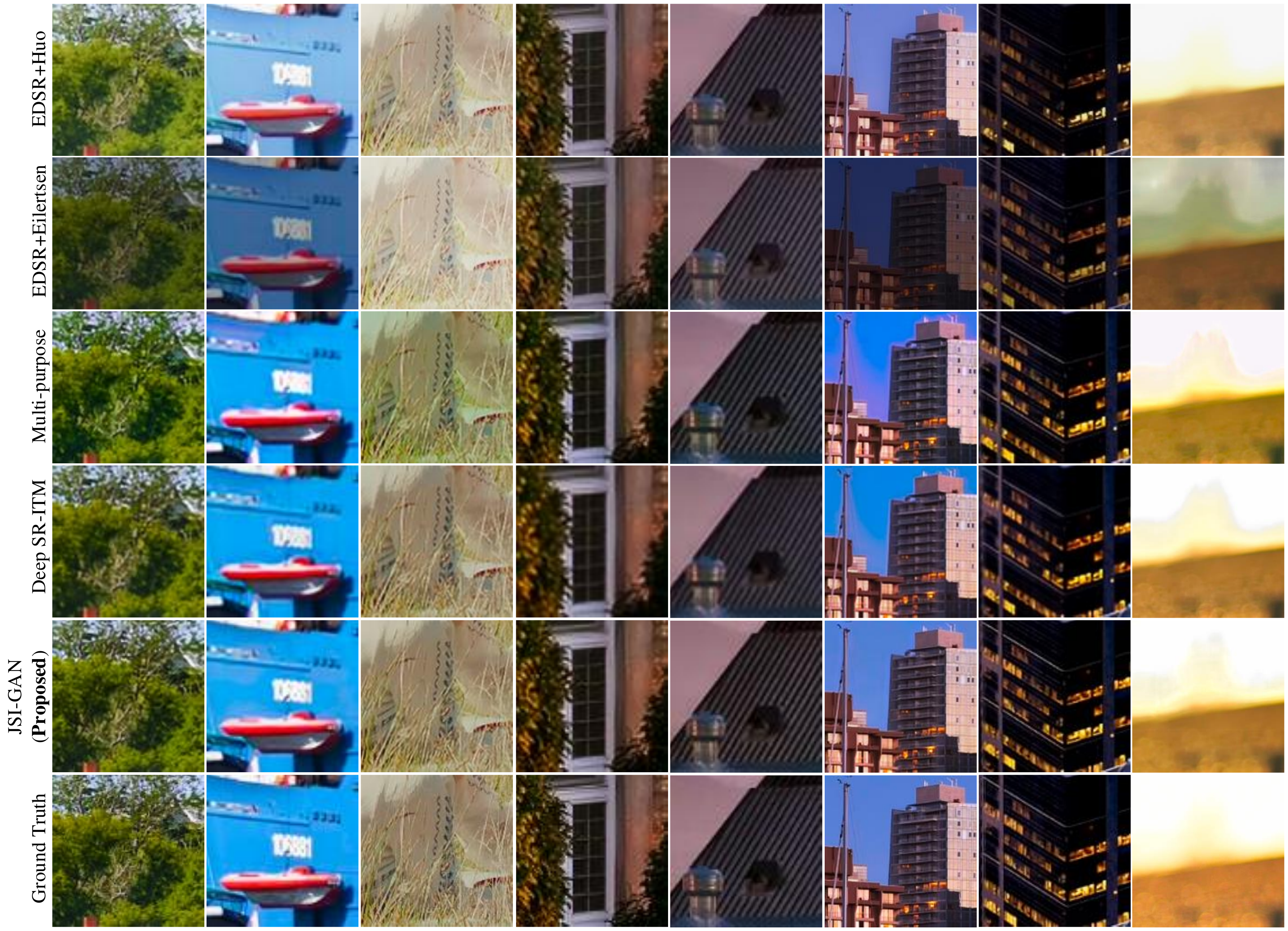}
\caption{Qualitative Comparison.}
\label{fig:5}
\end{figure*}

\subsubsection{Quantitative Comparison}
The quantitative comparison of the proposed JSInet and JSI-GAN against the previous methods is given in Table \ref{table:3}. The performance is measured using error-based metrics such as PSNR and mPSNR, and structural metrics such as SSIM and MS-SSIM, as well as HDR-VDP-2.2.1 \cite{hdr-vdp}, which is able to measure the performance degradation in all luminance conditions. The CNN-based joint SR-ITM methods outperform the cascaded methods by a large margin, and our JSInet outperforms the other methods in all cases except for HDR-VDP for scale factor 2. The proposed JSI-GAN also shows good quantitative performance, thanks to the stable training that mitigates the drop in objective performance metrics. 

Additionally, we provide the inference time in seconds and the number of parameters for the joint SR-ITM methods, which are denoted as \textit{Inf. Time} and \textit{No. Parameters} in Table \ref{table:3}, when converting to 4K target resolution measured with an NVIDIA TITAN Xp GPU. Note that only the generator parameters are computed for JSI-GAN, and the number of parameters increases proportionally with the scale factor due to the parameters dedicated for the local filters.

\subsubsection{Qualitative Comparison}
The qualitative comparison of the proposed JSI-GAN is given in Fig. \ref{fig:1} and Fig. \ref{fig:5}. In Fig. \ref{fig:1}, our method is able to reconstruct the fine lines on the window, produce more tree-like textures, and generate correct horizontal patterns on the wall. Likewise in Fig. \ref{fig:5}, JSI-GAN generates \textit{fine details} with \textit{enhanced contrast} without artifacts in the homogeneous regions, thanks to the task-specific subnets and the detail component-related losses.

\section{Conclusion}
In this paper, we first proposed a GAN-based framework for joint SR-ITM, called JSI-GAN, where the generator (JSInet) jointly optimizes the three task-specific subnets designed with pixel-wise 1D separable filters for fine detail restoration and pixel-wise 2D local filters for contrast enhancement. These subnets were carefully designed for their intended purposes to boost the output performance. Moreover, we also proposed a novel detail GAN loss alongside the conventional GAN loss, which helps enhancing both the local details and contrasts for generating high quality HR HDR reconstructions. We analyzed the efficacy of the subnet components and the weighting parameters for losses with intensive ablation studies. Our proposed JSI-GAN, which is applicable for directly converting LR SDR frames to HR HDR ones in real-world applications, achieves the state-of-the-art performance over the previous methods.

\section{Acknowledgement}
This work was supported by Institute for Information \& communications Technology Promotion (IITP) grant funded by the Korea government (MSIT) (No. 2017-0-00419, Intelligent High Realistic Visual Processing for Smart Broadcasting Media).

\end{document}